\documentclass[12pt]{article}
\usepackage{amsmath,amsfonts}
\usepackage{graphicx}
\usepackage{lscape}
\usepackage{multirow,tabularx}
\usepackage[sort]{natbib}
\usepackage{tikz}
\bibpunct{(}{)}{;}{a}{,}{,}

\def\Cov{\text{Cov}}
\def\Var{\text{Var}}

\begin{document}

\begin{center}
{\Large\bf MCMC Methods for Gaussian Process Models\\[-2pt]
           Using Fast Approximations for the Likelihood\\ }
\vspace{0.3in}

\begin{minipage}{0.45\linewidth}
\centering
Chunyi Wang\\
Department of Statistical Sciences\\
University of Toronto\\
\texttt{chunyi@utstat.toronto.edu}\\
~
\end{minipage}
\begin{minipage}{0.45\linewidth}
\centering
Radford M.\ Neal\\
Department of Statistical Sciences and \\
Department of Computer Science\\
University of Toronto\\
\texttt{radford@utstat.toronto.edu}
\end{minipage}
\\

\vspace{0.25in}

9 May 2013

\vspace{0.1in}

\end{center}

\begin{quotation}\noindent
Gaussian Process (GP) models are a powerful and flexible tool for
non-parametric regression and classification. Computation for GP
models is intensive, since computing the posterior density, $\pi$, for
covariance function parameters requires computation of the covariance
matrix, $C$, a $pn^2$ operation, where $p$ is the number of covariates
and $n$ is the number of training cases, and then inversion of $C$, an
$n^3$ operation.  We introduce MCMC methods based on the ``temporary
mapping and caching'' framework, using a fast approximation, $\pi^*$,
as the distribution needed to construct the temporary space. We
propose two implementations under this scheme: ``mapping to a
discretizing chain'', and ``mapping with tempered transitions'', both
of which are exactly correct MCMC methods for sampling $\pi$, even
though their transitions are constructed using an approximation.
These methods are equivalent when their tuning parameters are set at the
simplest values, but differ in general. We compare how well these
methods work when using several approximations, finding on 
synthetic datasets that a $\pi^*$ based on the ``Subset of Data''
(SOD) method is almost always more efficient than standard MCMC using
only $\pi$.  On some datasets, a more sophisticated $\pi^*$ based on
the ``Nystr\"om-Cholesky'' method works better than SOD.

\end{quotation}

\begin{section}{Introduction}

Evaluating the posterior probability density function is the most
costly operation when Markov Chain Monte Carlo (MCMC) is applied to
many Bayesian inference problems.  One example is the Gaussian Process
regression model (see Section \ref{sec:app} for a brief introduction),
for which the time required to evaluate the posterior probability density
increases with the cube of the sample size.  However, several fast but
approximate methods for Gaussian Process models have been developed.
We show in this paper how such an approximation to the posterior
distribution for parameters of the covariance function in a Gaussian
process model can be used to speed up sampling, using either of two
schemes, based on ``mapping to a discretizing chain'' or ``mapping
with tempered transitions''.  Both schemes produce an exactly correct
MCMC method, despite using an approximation to the posterior density
for some operations.

In the next section, we describe a general scheme for contructing
efficient MCMC methods using temporary mapping and caching techniques,
first introduced by \citet{Neal:2006}, which is the basis for both of
the schemes for using approximations that are introduced in this
paper.

One possibility for a space to temporarily map to is the space of
Markov chain realizations that leave a distribution $\pi^*$ invariant.
Our hope is that if we use such a space with a $\pi^*$ that is a good
approximation to $\pi$, but faster to compute, then MCMC with
temporary mapping and caching will be faster than MCMC methods using
only $\pi$.

We then consider how the tempered transiton method due to
\citet{Neal:1996} can also be viewed as mapping temporary to another
space.  Using this view, we give a different proof that detailed
balance holds for tempered transitions.  We then discuss how the
sequence of transitions
$\hat{T}_1,\hat{T}_2,...,\check{T}_2,\check{T}_1$ (which collectively
form the tempered transition) should be chosen when they are defined
using fast approximations, rather than (as in the original context for
tempered transtions) by modifying the original distribution, $\pi$, in
a way that does not reduce computation time.

We apply these two proposed schemes to Gaussian process regression
models that have a covariance function with unknown hyperparameters,
whose posterior distribution must be sampled using MCMC. We discuss
several fast GP approximation methods that can be used to contruct an
approximate $\pi^*$.  We conclude by presenting experiments on
synthetic datasets using the new methods that show that these methods
are indeed faster than standard methods using only $\pi$.

\end{section}

\begin{section}{MCMC with temporary mapping and caching} \label{sec:map-cache}

To start, we present two general ideas for improving MCMC ---
temporarily mapping to a different state space, and caching the
results of posterior density computations for possible later use.

\begin{subsection}{Creating Markov transitions using temporary mappings}
To obtain samples of a target distribution $\pi$ from space $\mathcal{X}$ 
using MCMC, we need to find a transition probability $T(x'|x)$, for which
\begin{equation}
        \label{eq:inv}
        \int \pi(x) T(x'|x) dx = \pi(x')
\end{equation}
i.e., $T(x'|x)$ leaves the target distribution $\pi$ invariant. There
are many ways to form such a transition. In the famous Metropolis
algorithm \citep{Metropolis:1953}, from a current state $x$, we
propose to move to a candidate state $x^*$ according to a 
proposal distribution $S(x'|x)$ that is symmetric (i.e., $S(x'|x)=S(x|x')$), 
and then accept this proposal with probability $\min(1,\pi(x^*)/\pi(x))$. If 
this proposal is accepted, the new state is $x'=x^*$, otherwise
$x'=x$. It's easy to show that these transitions leave $\pi$ invariant
(in fact they satisfy the stronger ``detailed balance'' condition that
$\pi(x)T(x'|x)=\pi(x')T(x|x')$).

The temporary mapping technique \citep{Neal:2006} defines such a
transition via three other stochastic mappings, $\hat{T}$, $\bar{T}$
and $\check{T}$, as follows:
\begin{equation}
        \label{eq:mapping}
        x \stackrel{\hat{T}}{\longrightarrow} y \stackrel{\bar{T}}{\longrightarrow} y' \stackrel{\check{T}}{\longrightarrow} x'
\end{equation}
where $x,x' \in \mathcal{X}$ and $y,y' \in \mathcal{Y}$. Starting from $x$, we obtain a value $y$ in the temporary space $\mathcal{Y}$ by $\hat{T}(y|x)$. The target distribution for $y$ has probability mass/density function $\rho(y)$. We require that
\begin{equation}
        \label{eq:mapup}
        \int \pi(x)\hat{T}(y|x) dx = \rho(y)
\end{equation}
We then obtain another sample $y'$ using $\bar{T}(y'|y)$, which leaves $\rho$ invariant:
\begin{equation}
        \label{eq:mapbar}
        \int \rho(y)\bar{T}(y'|y) dy = \rho(y')
\end{equation}
Finally, we map back to $x'\in \mathcal{X}$ using $\check{T}(x'|y)$, which we require to satisfy
\begin{equation}
        \label{eq:mapdown}
        \int \rho(y') \check{T}(x'|y') dy' = \pi(x')
\end{equation}

        It's easy to see that the combined transition $T(x'|x) = \int\int \hat{T}(y|x)\bar{T}(y'|y)\check{T}(x'|y') dydy'$ leaves $\pi$ invariant:
\begin{eqnarray}
\int \pi(x) T(x'|x) dx & = & \int\int\int \pi(x)\hat{T}(y|x)\bar{T}(y'|y)
                             \check{T}(x'|y') dydy'dx \\
  & = & \int\int \rho(y)\bar{T}(y'|y)\check{T}(x'|y')dydy' \\
  & = & \int \rho(y')\check{T}(x'|y')dy' \\
  & = & \pi(x')
\end{eqnarray}

Quite a few existing methods can be viewed as mapping to temporary
spaces. For instance, the technique of temporarily introducing
auxiliary variables can be considered as mapping from $x$ to
$y=(x,z)$, where $z$ is a set of auxiliary variables. 

\end{subsection}

\begin{subsection}{Caching values for future re-use}

Many MCMC transitions require evalulating the probability density of
$\pi$, up to a possibly unknown normalizing constant.  For example,
each iteration of the Metropolis algorithm needs the probability
density values of both the current state $x$ and the candidate state
$x^*$. Since these evaluations typically dominate the MCMC computation
time, it may be desirable to save (`cache') computed values of
$\pi(x)$ so they can be re-used when the same state $x$ appears in the
chain again.

Caching is always useful for the Metropolis algorithm, since if we
reject a proposal $x^*$, we will need $\pi(x)$ for the next
transition, and if we instead accept $x^*$ then it becomes the current
state and we will need $\pi(x^*)$ for the next transition. 

When the proposal distribution is discrete (as it will always be when
the state space is discrete), the probability of proposing an $x^*$
that was previously proposed can be positive, so saving the computed
value of $\pi(x^*)$ may be beneficial even if $x^*$ is rejected.  When
the state space is continuous, however, the proposal distributions
commonly used are also continuous, and we will have zero probability
of proposing the same $x^*$ again.  But in this case, as we will see
next, caching can still be beneficial if we first map to another space
with a ``discretizing chain''.  \end{subsection}

\end{section}

\begin{section}{Mapping to a discretizing chain}\label{sec:map}

To take full advantage of both mapping and caching, we propose a
temporary mapping scheme where the temporary space is continuous, but
is effectively discrete with regard to transitions $\bar{T}$.

        Let $R(x'|x)$ be the transition probabilities for a Markov Chain which leaves $\pi^*$ invariant. Let $\tilde{R}(x|x')=R(x'|x)\pi^*(x)/\pi^*(x')$ be the reverse transition probabilities, which clearly also leave $\pi^*$ invariant.

        We map from $\mathcal{X}$ to $\mathcal{Y}$, a space of realizations of this Markov Chain of length $K$, where one time step of this chain is ``marked''. To map $x\in \mathcal{X}$ to $y\in \mathcal{Y}$, we use a $\hat{T}$ that operates as follows:

        \begin{itemize}
                \item Choose $k$ uniformly from $0,...,K-1$.
                \item Simulate $K-1-k$ forward transition steps using $R$ starting at $x_k=x$, producing states $x_{k+1},...,x_{K-1}$.
                \item Simulate $k$ reverse transitions using $\tilde{R}$, starting at $x_k=x$, producing states $x_{k-1},...,x_0$.
                \item Set the ``marked'' time step to $k$.
        \end{itemize}
        
The transition $\bar{T}$ moves the mark along the chain from $k$ to
another time step $k' \in \{0,\ldots,K\!-\!1\}$, while keeping the current
chain realization, $(x_0,\ldots,x_{K-1})$, fixed. The transition
$\check{T}$ just takes the marked state, so $x'=x_{k'}$. The actual
implementation will not necessarily simulate all $K-1$ steps of the
discretizing chain --- a new step is simulated only when it is
needed. We can then let $K$ go to infinity, so that $\bar{T}$ can move
the mark any finite number of steps forward or backward.

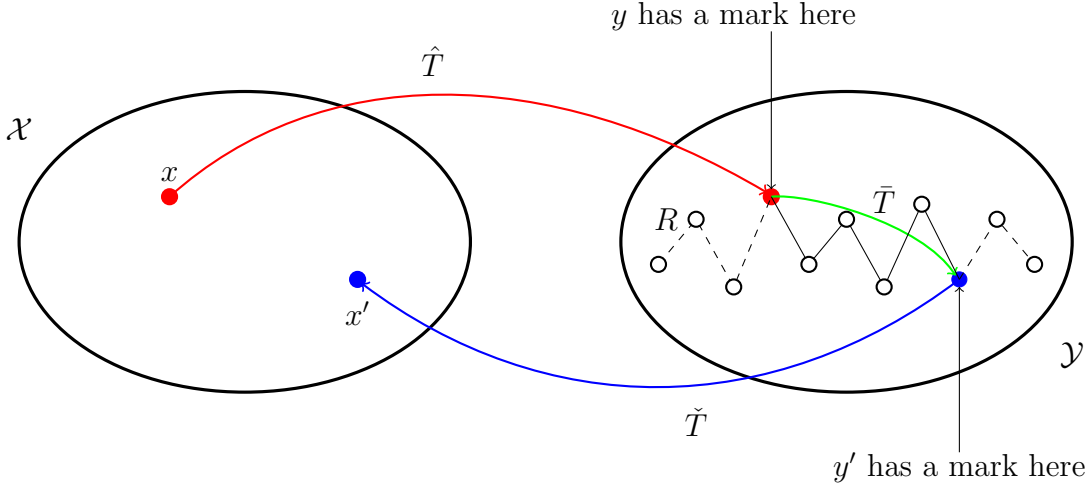
\begin{figure}[t]
  \centering
  \begin{tikzpicture}
                \draw[very thick] (-4,0) ellipse (3 and 2); \node[draw=none] at (-7,1.5) {$\mathcal{X}$}; 
                \draw[very thick] ( 4,0) ellipse (3 and 2); \node[draw=none] at (7,-1.5) {$\mathcal{Y}$}; 
                \draw[->,red, thick] (-5,0.6) .. controls (-3,2.4) and (0,2.4) ..  (2.95,0.65); \node[draw=none] at (-1.5,2.4) {$\hat{T}$};                         
                \draw[red,thick,fill=red] (-5,0.6) circle [radius=0.1];\draw[red,thick,fill=red] (3,0.6) circle [radius=0.1]; 
                \draw[->,blue, thick] (5.5,-0.5) .. controls (3,-2.4) and (0,-2.4) ..  (-2.45,-0.55); \node[draw=none] at (2,-2.4) {$\check{T}$};
                \draw[blue,thick,fill=blue] (-2.5,-0.5) circle [radius=0.1];\draw[blue,fill=blue] (5.5,-0.5) circle [radius=0.1];

                \draw[->,thick,green] (3,0.6) .. controls (3.5,0.65) and (5,0.25) .. (5.45,-0.45); \node[draw=none] at (4.5,0.55) {$\bar{T}$};
                \node[draw=none] at (-5,0.9) {$x$}; 
                \node[draw=none] at (2.5,3) {$y$ has a mark here}; \draw[->] (3,2.8) -- (3,0.7);
                \node[draw=none] at (5.5,-3) {$y'$ has a mark here}; \draw[->] (5.5,-2.8) -- (5.5,-0.6);
                \node[draw=none] at (-2.5,-0.95) {$x'$};
                \draw[dashed] (1.5,-0.3) -- (2,0.3) -- (2.5,-0.6) -- (3,0.6);
                \draw (3,0.6) -- (3.5,-0.3) -- (4,0.3) -- (4.5,-0.6) -- (5,0.5) -- (5.5,-0.5);
                \draw[dashed] (5.5,-0.5) -- (6,0.3) -- (6.5,-0.3);

                \draw[black,thick,fill=white] (1.5,-0.3) circle [radius=0.1]; 
                \draw[black,thick,fill=white] (2,0.3) circle [radius=0.1]; 
                \draw[black,thick,fill=white] (2.5,-0.6) circle [radius=0.1];         
                \draw[black,thick,fill=white] (3.5,-0.3) circle [radius=0.1]; 
                \draw[black,thick,fill=white] (4,0.3) circle [radius=0.1]; 
                \draw[black,thick,fill=white] (4.5,-0.6) circle [radius=0.1]; 
                \draw[black,thick,fill=white] (5,0.5) circle [radius=0.1]; 
                \draw[black,thick,fill=white] (6,0.3) circle [radius=0.1]; 
                \draw[black,thick,fill=white] (6.5,-0.3) circle [radius=0.1]; 

                \node[draw=none] at (1.6,0.3) {$R$};
  \end{tikzpicture}
        \caption{Mapping to a discretizing chain and back.}
        \label{fig:mapping}
\end{figure}

Figure \ref{fig:mapping} illustrates this scheme. Note that an element
$y \in \mathcal{Y}$ is a chain realization with a mark placed on the
time step $k$. We write $y=(k;x_0,...,x_{K-1})$. When we say we ``move
the mark from $k$ to $k'$'', we actually use a transition $\bar{T}$ to
move from $y=(k;x_0,...,x_{K-1})$ to $y'=(k';x_0,...,x_{K-1})$, where
$y$ and $y'$ share the same chain realization and differ only on the
marked position. We are free to choose the way $\bar{T}$ moves the
mark in any way that leaves $\rho$ invariance --- for instance, we can
pick a number $s$ and propose to move mark from $k$ to $k+s$ or $k-s$
with equal probabilities.  We can make $r$ such moves within each
mapping. The discretizing chain makes the state space effectively
discrete, even though the space $\mathcal{Y}$ is continuous, and
consequently, when we move the mark around the chain realization,
there is a positive probability of hitting a location that has been
visited before.

        The transition $\bar{T}$ has to leave $\rho(y)$ invariant. We compute the ratio of $\rho(y')$ and $\rho(y)$ to see how we can construct a such a $\bar{T}$. $\rho$ has been implicitly defined in \eqref{eq:mapup} as the distribution resulting from applying $\hat{T}$ to $x$ drawn from $\pi$. The probability to sample $y$ is given by the simulation process described above (i.e. start from $x$, simulate $K-1-k$ forward steps using $R$ and $k$ backward steps using $\tilde{R}$), namely, if $y=(k;x_0,...,x_{K-1})$,
        \begin{align}\notag
                \rho(y) &= \pi(x_k)\frac{1}{K}R(x_{k+1}|x_k)\cdots R(x_{K-1}|x_{K-2})\times \tilde{R}(x_{k-1}|x_k)\cdots \tilde{R}(x_0|x_1)\\[3pt]
                \label{eq:rho_k}
                &=\frac{\pi(x_k)}{\pi^*(x_k)}\frac{1}{K} \underbrace{\pi^*(x_k)R(x_{k+1}|x_k)\cdots R(x_{K-1}|x_{K-2})\times \tilde{R}(x_{k-1}|x_k)\cdots \tilde{R}(x_0|x_1)}_{:=A}
        \end{align}
        An expression for $\rho(y')$ can be similarly obtained for $y'=(k';x_0,...,x_{K-1})$:
        \begin{align}
                \label{eq:rho_k'}
                \rho(y') &=\frac{\pi(x_{k'})}{\pi^*(x_{k'})}\frac{1}{K} \underbrace{\pi^*(x_{k'})R(x_{k'+1}|x')\cdots R(x_{K-1}|x_{K-2})\times \tilde{R}(x_{k'-1}|x_{k'})\cdots \tilde{R}(x_0|x_1)}_{:=A'}
        \end{align}
        We take out a factor of the ratio of densities $\pi/\pi^*$ from both \eqref{eq:rho_k} and \eqref{eq:rho_k'}, and write the remaining term as $A$ or $A'$, as indicated in the respective equation. Since $R$ and $\tilde{R}$ are reverse transitions with respect to $\pi^*$, if $k'>k$, then
        \begin{align} \notag
                \lefteqn{\pi^*(x_k)R(x_{k+1}|x_k)\cdots R(x_{k'}|x_{k'-1})} \\
                \notag \ \
                &\quad = \tilde{R}(x_k|x_{k+1}) \pi^*(x_{k+1})R(x_{k+2}|x_{k+1})\cdots R(x_{k'}|x_{k'-1}) \\  \notag
                &\qquad \vdots \\
                &\quad=\tilde{R}(x_k|x_{k+1})...\tilde{R}(x_{k'-1}|x_{k'}) \pi^*(x_{k'})
        \end{align}
It therefore follows that $A=A'$. A similar argument shows that $A=A'$ when
$k' \le k$. Thus the ratio of $\rho(y')$ and $\rho(y)$ is
        \begin{align}
                \frac{\rho(y')}{\rho(y)} 
                \label{eq:rho_kk'}
                &=\frac{\pi(x_{k'})/\pi^*(x_{k'})}{\pi(x_k)/\pi^*(x_k)}
        \end{align}
        Equation \eqref{eq:rho_kk'} implies that to leave $\rho$ invariant we can use a Metropolis type transition, $\bar{T}$, that proposes to move the mark from $k$ to $k'$ and accepts the move with probability
        \[ \min\left(1,\frac{\pi(x_{k'})/\pi^*(x_{k'})}{\pi(x_k)/\pi^*(x_k)}\right)\]
Note that if $\pi=\pi^*$, then the transition $\bar{T}$ will accept a
move of the mark to any other time step on the discretizing chain,
since the discretizing chain actually leaves the target
distribution $\pi^*$ invariant and therefore every time step of this
chain is a valid sample of $\pi$. If $\pi^*\ne \pi$, but is very
similar to $\pi$, we can hope the acceptance rate will be high. In
addition, if the evaluation of $\pi^*(x)$ takes much less time than that
of $\pi(x)$, mapping to the discretizing chain and then proposing 
large moves of the mark can save computation time,
since it effectively replaces evaluations of $\pi$ with evaluations of $\pi^*$,
except for the acceptance decisions..
On the other hand, if $\pi^*$
is completely arbitrary, the acceptance rate will be low, and if the
evalution of $\pi^*$ is not much faster than $\pi(x)$, we will not save
computation time. These $\pi^*$'s are not useful. We need 
$\pi^*$ to be a fast but good approximation to $\pi$. We will discuss this 
in the context of GP models in a later section.

        Every time we map into a temporary space, we can make multiple $\bar{T}$ updates (move the ``mark'' several times). This way we can take advantage of the ``caching'' idea, since sometimes the mark will be moved to a state where $\pi$ has already been computed, and therefore no new computation is needed. The number of updates is a tuning parameter, which we denote as ``$r$''. Another tuning parameter, which we denote as ``$s$'', is the number of steps of transition $R$ to ``jump'' when we try to move the mark. Note that although we only ``bring back'' (using $\check{T}$) the last updated sample as $x'$, all of the marked states are valid samples of $\pi(x)$, and can be used for computing expectations with respect to $\pi$ if desired.

\end{section}

\begin{section}{Tempered transitions} \label{sec:tmp}

The ``tempered transitions'' method of \citet{Neal:1996} can also be
viewed as mapping to a temporary space.  This method aims to sample
from $\pi$ using a sequence of distributions $\pi=\pi_0,\ \pi_1,
\ldots,\ \pi_n$.

For $i=0,\ldots,n$, let $\hat{T}_i$ (called the ``up'' transition) and
$\check{T}_i$ (the ``down'' transition) be mutually reversible
transitions with respect to the density $\pi_i$ --- i.e. for any pair of
states $x_i$ and $x_i'$,
        \begin{equation}
                \pi_i(x_i)\hat{T}_i(x_i'|x_i) = \check{T}_i(x_i|x_i') \pi_i(x_i') 
                \label{eq:mutual_reversibility}
        \end{equation}
This condition implies that both $\hat{T}_i$ and $\check{T}_i$ have
$\pi_i$ as their invariant distribution. If $\hat{T}_i = \check{T}_i$
then \eqref{eq:mutual_reversibility} reduces to the detailed balance
condition. If $\hat{T}_i=S_1S_2...S_k$ with all of $S_i$ being
reversible transitions, then $\check{T}_i=S_kS_{k-1}...S_1$ would
satisfy condition \eqref{eq:mutual_reversibility}.

        We map from $x\in\mathcal{X}$ to $y\in\mathcal{Y}$, a space of realizations of tempered transitions, using a $\hat{T}$ that operates as follows:
\begin{quotation}
        Generate $\hat{x}_1$ from $x$ using $\hat{T}_1$;

        Generate $\hat{x}_2$ from $\hat{x}_1$ using $\hat{T}_2$;

        \indent\indent\vdots

        Generate $\bar{x}_n$ from $\hat{x}_{n-1}$ using $\hat{T}_n$.

        Generate $\check{x}_{n-1}$ from $\bar{x}_n$ using $\check{T}_n$;

        Generate $\check{x}_{n-2}$ from $\check{x}_{n-1}$ using $\check{T}_{n-1}$;

        \indent\indent\vdots
        
        Generate $x^*$ from $\check{x}_1$ using $\check{T}_1$.
\end{quotation}
An element $y\in\mathcal{Y}$ can be written as $y=(x,\hat{x}_1,...,\bar{x}_n,...,\check{x}_1,x^*)$. 

$\bar{T}$ attempts to flip the order of $y$, accepting the flip with probability
        \begin{equation}
                \min\left(1,
                        \frac{\pi_1(\hat{x}_0)}{\pi_0(\hat{x}_0)}\cdots\frac{\pi_n(\hat{x}_{n-1})}{\pi_{n-1}(\hat{x}_{n-1})}\cdot
                        \frac{\pi_{n-1}(\check{x}_{n-1})}{\pi_n(\check{x}_{n-1})}\cdots\frac{\pi_0(\check{x}_0)}{\pi_1(\check{x}_0)}
                        \right)
                \label{eq:temper_acc_prob}
        \end{equation}                
where $\hat{x}_0$ and $\check{x}_0$ are synonyms for
$x$ and $x^*$, respectively, to keep notations consistent.
In other words, with this probability, we set $y'$ to
$y^*=(x^*,\check{x}_1,...,\bar{x}_n,...,\hat{x}_1,x)$ (the order is
reversed); otherwise we sset $y'=y$ (the order is preserved).

Finally, $\check{T}$ maps back to $x'\in\mathcal{X}$ by taking the
first coordinate of $y'$ (either the original $x$ or $x^*$, depending
on whether or not the flip was accepted).

Using the temporary mapping perspective, we can show that tempered
transitions are valid updates, leaving $\pi$ invariant, by defining
$\rho$ to be the result of applying $\hat T$ to a point drawn from
$\pi$, and then showing that $\bar T$ leaves $\rho$ invariant, and
that $\check T$ produces a point distributed as $\pi$ from a point
distributed as $\rho$.

The $\hat T$ mapping from $x=\hat{x}_0$ to
$y=(\hat{x}_0,\hat{x}_1,...,\bar{x}_n,...,\check{x}_1,\check{x}_0)$
involves a sequence of transitions:
        \[ \hat{x}_0\stackrel{\hat{T}_1}{\longrightarrow}\hat{x}_1\stackrel{\hat{T}_2}{\longrightarrow}\hat{x}_2\longrightarrow\cdots\longrightarrow\hat{x}_{n-1}\stackrel{\hat{T}_n}{\longrightarrow}\bar{x}_n \stackrel{\check{T}_n}{\longrightarrow}\check{x}_{n-1}\stackrel{\check{T}_{n-1}}{\longrightarrow}\check{x}_{n-2}\longrightarrow\cdots\longrightarrow\check{x}_1\stackrel{\check{T}_1}{\longrightarrow}\check{x}_0 \]
The probability density, $\rho$, for $y$ can be computed from this as
        \begin{equation}
                        \rho(y) = \pi_0(\hat{x}_0)\hat{T}_1(\hat{x}_1|\hat{x}_0)\cdots\hat{T}_n(\bar{x}_n|\hat{x}_{n-1})
                                                \check{T}_n(\check{x}_{n-1}|\bar{x}_n)\cdots\check{T}_1(\check{x}_0|\check{x}_1) \label{eq:rhodef}
        \end{equation}
Similarly, 
        \begin{equation}
            \rho(y^*) = \pi_0(\check{x}_0)\hat{T}_1(\check{x}_1|\check{x}_0)\cdots\hat{T}_n(\bar{x}_n|\check{x}_{n-1})
                                                \check{T}_n(\hat{x}_{n-1}|\bar{x}_n)\cdots\check{T}_1(\hat{x}_0|\hat{x}_1)
        \end{equation}
        
Now we compute the ratio of probability densities of $y^*$ and $y$:
        \begin{align}
\notag
                \frac{\rho(y^*)}{\rho(y)} &= \frac{\pi_0(\check{x}_0)\hat{T}_1(\check{x}_1|\check{x}_0)\cdots\hat{T}_n(\bar{x}_n|\check{x}_{n-1}) \check{T}_n(\hat{x}_{n-1}|\bar{x}_n)\cdots\check{T}_1(\hat{x}_0|\hat{x}_1)}
                        { \pi_0(\hat{x}_0)\hat{T}_1(\hat{x}_1|\hat{x}_0)\cdots\hat{T}_n(\bar{x}_n|\hat{x}_{n-1})
                                                \check{T}_n(\check{x}_{n-1}|\bar{x}_n)\cdots\check{T}_1(\check{x}_0|\check{x}_1)} \\ \label{eq:reo1}
&=\pi_0(\check{x}_0)\cdot
\frac{\hat{T}_1(\check{x}_1|\check{x}_0)}{\check{T}_1(\check{x}_0|\check{x}_1)}
\cdots
\frac{\hat{T}_n(\bar{x}_n|\check{x}_{n-1}) \check{T}_n(\hat{x}_{n-1}|\bar{x}_n)}
                        {\check{T}_n(\check{x}_{n-1}|\bar{x}_n)\hat{T}_n(\bar{x}_n|\hat{x}_{n-1})}
\cdots
\frac{\check{T}_1(\hat{x}_0|\hat{x}_1)}{\hat{T}_1(\hat{x}_1|\hat{x}_0)} 
\cdot 
\frac{1}{\pi_0(\hat{x}_0)}\\
\label{eq:mut_rev}
&=\pi_0(\check{x}_0)\cdot
\frac{\pi_1(\check{x}_1)}{\pi_1(\check{x}_0)}\cdots
\frac{\pi_n(\bar{x}_n)}{\pi_n(\check{x}_{n-1})}\cdot 
\frac{\pi_n(\hat{x}_{n-1})}{\pi_n(\bar{x}_n)}\cdots
\frac{\pi_1(\hat{x}_0)}{\pi_1(\hat{x}_1)}\cdot
\frac{1}{\pi_0(\hat{x}_0)}
\\\label{eq:reo2}
&=\frac{\pi_1(\hat{x}_0)}{\pi_0(\hat{x}_0)}\cdots\frac{\pi_n(\hat{x}_{n-1})}{\pi_{n-1}(\hat{x}_{n-1})}\cdot
                        \frac{\pi_{n-1}(\check{x}_{n-1})}{\pi_n(\check{x}_{n-1})}\cdots\frac{\pi_0(\check{x}_0)}{\pi_1(\check{x}_0)}
        \end{align}
We obtain \eqref{eq:mut_rev} from the mutual reversibility property of the
transitions $\hat T_i$ and $\check T_i$,
and \eqref{eq:reo1} and \eqref{eq:reo2} simply by reordering terms.

From (\ref{eq:reo2}), we see that the probability of accepting the
flip from $y$ to $y^*$ given by (\ref{eq:temper_acc_prob}) is equal to
$\min(1,\rho(y^*)/\rho(y))$, and thus $\bar T$ satisfies detailed
balance with respect to $\rho$.  It is also clear from (\ref{eq:rhodef})
that the marginal distribution under $\rho$ of the first component of $y$
is $\pi_0=\pi$, and thus $\check T$ maps from $\rho$ to $\pi$.

The original motivation of the tempered transition method described by
\citet{Neal:2006} is to move between isolated modes of multimodal
distributions. The distributions $\pi_1,...,\pi_n$ are typically of
the same class as $\pi$, but broader, making it easier to move between
modes of $\pi$ (typically, as $i$ gets larger, the distribution
$\pi_i$ gets broader, thus making it more likely that modes have
substantial overlap).  Evaluating the densities for $\pi_1,...,\pi_n$
typically takes similar computation time as evaluating the density for
$\pi$. Our mapping-caching scheme, on the other hand, is designed to
reduce computation. Ideally, in our scheme the bigger $i$ is, the
faster is the evaluation of $\pi_i(x)$. One possibility for this is
that each $\pi_i$ is an approximation of $\pi$, and as $i$ increases
the computation of $\pi_i$ becomes cheaper (but worse). 

The two methods we propose in this paper are equivalent if
the following are all true:
        \begin{itemize}
                \item For mapping to a discretizing chain:
                \begin{enumerate}
                        \item The transition $R$ which leaves $\pi^*$ invariant is reversible. 
                        \item $s=2k$, i.e. $\bar{T}$ always attempts to move the mark over an even number of $R$ updates.                        
                        \item $r=1$, i.e. $\bar{T}$ attempts to move the mark only once within each mapping.
                \end{enumerate}
                \item For mapping by tempered transitions: 
                \begin{enumerate}
                        \item $n=1$, i.e., there is only one additional distribution.
                        \item $\hat{T}_1 = \check{T}_1=R^{k}$, i.e. 
  these transitions consist of $k$ updates using $R$ (and hence $\pi_1=\pi^*$).
                \end{enumerate}
        \end{itemize}

When all above are true except that $n>1$, so more than one additional
distribution is used in the tempered transitions, we might expect
tempered transitions to perform better, as they propose a new point
through the guidance of these additional distributions, and
computations for these additional distributions should be negligible,
if they are faster and faster approximations. On the other hand, we
might think that $r>1$ will improve the performance when mapping to a
discretizing chain, since then caching could be exploited. 
So each method may have its own advantages.

\end{section}

\begin{section}{Application to Gaussian process models}\label{sec:app}

We now show how these MCMC methods can be applied to Bayesian inference
for Gaussian process models.

\begin{subsection}{Introduction to Gaussian process models}

We start with a brief introduction to Gaussian process (GP) models to establish
notation.  The problem is to model the
association between covariates $x$ and a response $y$ using $n$
observed pairs $(x_1,y_1),...,(x_n,y_n)$, and then make predictions for
the $y$ in future items once their covariates, $x$, have been observed.
We can write such a model as
        \begin{equation}
                \label{reg}
                y_i = f(x_i) + \epsilon_i
        \end{equation}
where $x_i$ is a covariate vector of length $p$, and $y_i$ is the correspoding
scalar response. The $\epsilon_i$ are random residuals,
assumed to have Gaussian distributions with mean 0 and constant
variance $\sigma^2$.
 
Bayesian GP models assume that the noise-free function $f$ comes from
a Gaussian Process which has prior mean function zero and some
specified covariance function. Note that a zero mean prior is not a
requirement --- we could specify a non-zero prior mean function $m(x)$
if we have \textit{a priori} knowledge of the mean structure. Using a
zero mean prior just reflects prior knowledge that the function is
equally likely to be positive or negative; the posterior mean of the
function is typically not zero.

The covariance function could be fixed \textit{a priori}, but more
commonly is specified in terms of unknown hyperparameters, $\theta$, which are
then estimated from the data. Given the values of the hyperparameters,
the response $y$ follows a multivariate Gaussian distribution with
zero mean and a covariance matrix given by
        \begin{align}
                \label{covy}
                \Cov(y_i,y_j) &\ =\ K(x_i,x_j) + \Cov(\epsilon_i,\epsilon_j)
                 \ =\ K(x_i,x_j)  + \delta_{ij}\sigma^2
        \end{align}
where $\delta_{ii} = 1$ and $\delta_{ij}=0$ when $i\ne j$, and
$K$ is the covariance function of $f$. Any covariance function that
always leads to a positive semi-definite covariance matrix can be used.
One example is the squared exponential
covariance function with isotropic length-scale (to which we add a constant
allowing the overall level of the function to be shifted from zero):
        \begin{equation}
                \label{eq:covSEiso}
                K(x_i,x_j) = c^2 +  \eta^2 \exp\left( -\frac{\|x_{i}-x_{j}\|^2}{\rho^2}\right)
        \end{equation}
Here, $c$ is a fairly large constant (not excessively large, to avoid
numerical singularity), and $\eta$, $\sigma$, and $\rho$ are
hyperparameters --- $\eta$ controls the magnitude of variation of $f$,
$\sigma$ is the residual standard deviation, and $\rho$ is a length scale 
parameter for the covariates. We can instead assign a different length scale 
to each covariate, which leads to the squared exponential covariance
function with automatic relevance determination (ARD):
        \begin{equation}
                \label{covfun}
                K(x_i,x_j) = c^2 +  \eta^2 \exp\left( -\sum_{k=1}^p \frac{(x_{ik}-x_{jk})^2}{\rho_k^2}\right)
        \end{equation}
Unless noted otherwise,  we will use the squared exponential covariance functions \eqref{eq:covSEiso} or \eqref{covfun} thoughout this paper.

When the values of the hyperparameters are known, the predictive
distribution for the response, $y_*$, a test case with covariates
$x_*$, based on observed values $x=(x_1,...,x_n)$ and $(y_1,...,y_n)$, 
is Gaussian with the following mean and variance:
        \begin{equation}
                \label{predmean}
                E(y_*|x,y,x_*,\theta) = k^TC(\theta)^{-1}y
        \end{equation}
        \begin{equation}
                \label{predvar}
                \Var(y_*|x,y,x_*,\theta) = v - k^TC(\theta)^{-1}k
        \end{equation}
In the equations above, $k$ is the vector of covariances between $y_*$
and each of $y_i$, $C(\theta)$ is the covariance matrix of the
observed $y$, based on the known hyperparameters $\theta$, and $v$ is
the prior variance of $y_*$, which is $\Cov(y_*,y_*$) from \eqref{covy}.

When the values of the hyperparameters are
unknown, and therefore must be estimated from the data, we put a
prior, $p(\theta)$, on them (typically an independent Gaussian prior
on the logarithm of each hyper-parameter), and obtain the posterior
distribution $p(\theta|x,y) \propto \mathcal{N}(y|0,C(\theta))\,p(\theta)$. 
The predictive mean of $y$ is then computed by integrating over the posterior 
distribution of the hyperparameters:
        \begin{equation}
                \label{predmeanint}
 E(y_*|x,y,x_*) = \int_{\Theta} k^TC(\theta)^{-1}y \cdot p(\theta|x,y)\, d\theta
        \end{equation}
The predicted variance is given by
        \begin{align}
                \label{predvarint}
\mbox{Var}(y_*|x,y,x_*) & \ = \
E[\mbox{Var}(y_*|x,y,x_*,\theta)\,|\,x,y] \ + \
\mbox{Var}[E(y_*|x,y,x_*,\theta)\,|\,x,y] 
        \end{align}

Finding $C^{-1}$ directly takes time proportional to $n^3$, but we do
not have to find the inverse of $C$ explicitly. Instead we find the
Cholesky decomposition of $C$, denoted as $R=\mbox{chol}(C)$, for
which $R^TR=C$ and $R$ is an ``upper'' triangular matrix (also called 
a ``right'' triangular matrix). This also takes time proportional to $n^3$,
but with a much smaller constant.  We then solve $R^Tu = y$ for $u$
using a series of forward subsititutions (taking time proportional to $n^2$).  
From $R$ and $u$, we can compute the likelihood for $\theta$, which is
needed to compute the posterior density, by making use of the expressions
\begin{equation}
                y^TC^{-1} y = y^T(R^TR)^{-1}y = y^TR^{-1} \left(R^{T}\right)^{-1}y = u^T u
\end{equation}
and
\begin{equation}
                \det(C) = \det(R)^2 = \prod_{i=1}^n R_{ii}^2 
\end{equation}
Similarly, equations (\ref{predmean}) and (\ref{predvar}) and be reformulated
to use $R$ rather than $C^{-1}$.

\end{subsection}

\begin{subsection}{Approximating $\pi$ for GP models}

As discussed in Section \ref{sec:map}, using a poor $\pi^*$ for the
discretizing chains on $\mathcal{Y}$, or poor $\pi_i$ for tempered
transitions, can lead to a poor MCMC method which is not useful.  We
would like to choose approximations to $\pi$ that are good, but that can
nevertheless be computated much faster than $\pi$.  For GP regression
models, $\pi$ will be the posterior distribution of the
hyperparameters, $\theta$.

Quite a few efficient approximation methods for GP models have been
discussed from a different perspective. For example, \citet{QC:2007}
categorizes these approximations in terms of ``effective prior''. Most
of these methods are used for approximate training and prediction;
not all of them are suitable for forming a posterior approximation, $\pi^*$. 
For example, we cannot take advantage of an efficient approximated prediction. 

\begin{subsubsection}{Subset of data (SOD)}

The most obvious approximation is to simply take a
subset of size $m$ from the $n$ observed pairs $(x_i,y_i)$ and use the
posterior distribution given only these observations as $\pi^*$:
        \begin{align}
          \pi^*(\theta) &\ =\ \mathcal{N}(y|0,\hat{C}_{(m)}(\theta))\,p(\theta) 
        \label{eq:pistar}
        \end{align}
where $p(\theta)$ is the prior for $\theta$, the vector of
hyperparameters, and $\mathcal{N}(a|\mu,\Sigma)$ denotes the
probability density of a multivariate normal distribution
$N(\mu,\Sigma)$ evaluated at $a$. $\hat{C}_{(m)}(\theta)$ is computed based on
hyperparameters $\theta$ and the $m$ observations in the subset.

Even though the SOD method seems quite naive, it does speed up
computation of the Cholesky decomposition of $C$ from time
proportional to $n^3$ to time proportional to $m^3$.
If a small subset (say 10\% of the full dataset) is used to form
$\pi^*$, we can afford to do a lot of Markov chain updates for $\pi^*$,
since the time it takes to make these updates will be quite small
compared to a computation of $\pi$. 
So a $\pi^*$ formed by this method might still be useful.

To form a $\pi^*$ using SOD, we need the following major computations, if
there are $p$ covariates:\\

        \begin{tabular}{c|c}
                \hline
                Operation & Complexity \\ \hline
                Compute $\hat{C}_{(m)}$ & $pm^2$ \\ \hline
                Find chol($\hat{C}_{(m)}$) & $m^3$ \\ \hline
        \end{tabular}

\vspace{6pt}

\end{subsubsection}
\begin{subsubsection}{Using low-rank plus diagonal matrices}

A covariance matrix in a GP model typically has the form $C \,=\,K + \sigma^2I$,
where $K$ is the noise-free covariance matrix, and $\sigma^2$ is the
residual variance.  More generally, if the residual variance differs for
different observations, the covariance matrix will be $K$ plus a diagonal
matrix giving these residual variances.  If we approximate $K$ 
by a matrix $\hat{K}$ with rank $m<n$, and let $\hat{C} = \hat{K}+\sigma^2I$,
then after writing $\hat{K} = B S B^T$, where $B$ is $n$ by $m$, we can
quickly find $\hat{C}^{-1}$ by taking advantage of the 
matrix inversion lemma, which states that
  \begin{equation}
    (BSB^{T}+D)^{-1} = D^{-1} - D^{-1}B(S^{-1}+B^TD^{-1}B)^{-1}B^TD^{-1}
    \label{eq:minv}
  \end{equation}
This can be simplified as follows when $D=dI$, where $d$ is a scalar, 
$B$ has orthonormal columns (so that $B^T B=I$), and $S$ is a diagonal matrix with
diagonal elements given by the vector $s$, denoted by $\text{diag}(s)$:
\begin{align}
    (B\,\text{diag}(s)\,B^T+dI)^{-1}
    &= d^{-1}I - d^{-1}I B (\text{diag}(s^{-1}) + B^Td^{-1}IB)^{-1} B^T d^{-1} I\\
    &= d^{-1}I - d^{-2} B(\text{diag}(1/s)+ B^TB/d)^{-1}B^T \\
    &= d^{-1}I - d^{-1} B(\text{diag}(d/s)+ I)^{-1} B^T \\
    &= d^{-1}I - d^{-1} B(\text{diag}((s+d)/s)))^{-1}B \\
    &= d^{-1}I - B\,\text{diag}(s/(d(s+d)))\,B^T \label{eq:minvd}
\end{align}
Expressions above such as $1/s$ denote element-by-element arithmetic
on the vector operands.


We can use the matrix determinant lemma to compute the determinant of $\hat{C}$.
  \begin{equation}
    \text{det}(BSB^T+D) \ =\ \text{det}(S^{-1}+B^TD^{-1}B)\,\text{det}(D)\,\text{det}(S)
  \end{equation}
When $D=dI$ with $d$ being a scalar, $\text{det}(D)=d^n$ is
trivial, and $\text{det}(S^{-1}+B^TD^{-1}B)$ can be
found from the Cholesky decomposition of $S^{-1}+B^TD^{-1}B$.

  Once we obtain $\hat{C}^{-1}$ and $\text{det}(\hat{C})$, we can easily establish our $\pi^*$:
  \begin{equation}
    \pi^*(\theta) = \mathcal{N}(y|0,\hat{C})p(\theta)
  \end{equation}

\end{subsubsection}

\begin{subsubsection}{The Eigen-exact approximation}

Since the noise-free covariance matrix, $K$, is non-negative definite,
we can write it as $K = E\Lambda E^T=\sum_i^n \lambda_i e_i e_i^T$,
where $E$ has columns $e_1,e_2,...,e_n$, the eigenvectors of $K$, and
the diagonal matrix $\Lambda$ has the eigenvalues of $K$,
$\lambda_1\ge \lambda_2 \ge ...\ge \lambda_n$ on its diagonal. This is
known as the eigendecomposition. A natural choice of low-rank plus
diagonal approximation would be $\hat{C} = \hat{K} + \sigma^2I$ where
$\hat{K} = BSB^T$ where $B$ is an $n\times m$ matrix with columns
$e_1,...,e_m$, and $S$ is a diagonal matrix with diagonal entries
$\lambda_1,...,\lambda_m$.  We expect this to be a good approximation
if $\lambda_{m+1}$ is close to zero.

With this approximation, $\hat{C}^{-1}$ can be computed rapidly from
$B$ and $S$ using (\ref{eq:minvd}). However, the time needed to find
the first $m$ eigenvalues and eigenvectors (and hence $B$ and $S$) is
proportional to $mn^2$, with a much larger constant factor than for
the $n^3$ computation of all eigenvalues and eigenvectors. In
practice, depending on the values of $m$ and $n$ and the software
implementation, a $\pi^*$ formed by this method could even be slower
than the original $\pi$.  Since our experiments confirm this, we
mention it here only because it is a natural reference point.

\end{subsubsection}

\begin{subsubsection}{The Nytr\"om-Cholesky approximation}

In the Nystr\"om method, we take a random $m$ by $m$ submatrix of the
noise-free covariance matrix, $K$, which is equivalent to looking at the
noise-free covariance for a subset of the data of size $m$, and then
find its eigenvalues and eigenvectors.  This takes time proportional
to $m^3$.  We will denote the submatrix chosen by $K^{(m,m)}$, and its
eigenvalues and eigenvectors by $\lambda_1^{(m)},...,\lambda_m^{(m)}$
and $e_1^{(m)},...,e_m^{(m)}$.
We can then approximate the first $m$ eigenvalues and eigenvectors of the 
full noise-free covariance matrix by\vspace{-10pt}
        \begin{align}
                \hat{\lambda}_i &= (n/m)\lambda_i^{(m)} \\
                \hat{e}_i &= \frac{\sqrt{m/n}}{\lambda_i^{(m)}} K^{(n,m)} e_i^{(m)}
        \end{align}
where $K^{(n,m)}$ is the $n$ by $m$ submatrix of $K$ with only the columns
corresponding to the $m$ cases in the random subset.

The covariance matrix $C$ can then be approximated in the same fashion
as Eigen-exact, with the exact eigenvalues and eigenvectors replaced
by the approximated eigenvalues $\hat{\lambda_1},...,\hat{\lambda}_m$
and eigenvectors $\hat{e}_1,...\hat{e}_m$. However, a more efficient
computational method for this approximation, requiring no eigenvalue/eigenvector
computations, is available as follows:
        \begin{equation}
                \hat{K} \,=\, K^{(n,m)} [K^{(m,m)}]^{-1} K^{(m,n)}
                \label{eq:nystrom-cholesky}
        \end{equation}
where $K^{(m,n)} = [K^{(n,m)}]^T$).
We can find the
Cholesky decomposition of $K^{(m,m)}$ as $R^TR$, in time proportional to $m^3$,
with a much smaller constant factor than finding the
eigenvalues and eigenvectors. Equation \eqref{eq:nystrom-cholesky} can then be
put in the form of $BSB^T$ by letting $B=K^{(n,m)}R^{-1}$ and $S=I$.
In practice, the noise free submatrix $K^{(m,m)}$ often has some
very small positive eigenvalues, which can appear to be negative due
to round-off error, making the Cholesky decomposition fail, a problem that
can be avoided by adding a small jitter to the diagonal \citep{Neal:1993}.

An alternative way of justifying the approximation in
(\ref{eq:nystrom-cholesky}) is by considering the covariance matrix
for the predictive distribution of all $n$ noise-free observations
from the random subset of $m$ noise-free observations, which (from a
generalization of (\ref{predvar})) is $K-K^{(n,m)} [K^{(m,m)}]^{-1}
K^{(m,n)}$.  When this is close to zero (so these $m$ noise-free
observations are enough to almost determine the function),
$\hat{K}$ will be almost the same as $K$.

More sophisticated schemes for Nystr\"om-Cholesky have been
proposed. For instance, \citet{Drineas:2005} randomly select the $m$
columns to construct $\hat{C}$ according to some
``judiciously-chosen\rq{}\rq{} and data-dependent probability
distribution rather than uniformly choose the $m$ columns. 

To form a $\pi^*$ using Nystr\"om-Cholesky, we need the following
major computations:\vspace{10pt}

        \begin{tabular}{c|c}
                \hline
                Operation & Complexity \\\hline
                Compute $K^{(n,m)}$ & $pmn$ \\\hline
                Find chol($K^{(m,m)}$) & $m^3$ \\\hline
        \end{tabular}

\vspace{4pt}

\end{subsubsection}

\end{subsection}
\end{section}

\begin{section}{Experiments} \label{sec:exp}

Here we report tests of the performance of the methods described in this paper 
using synthetic datasets. 

\begin{subsection}{Experimental setup}

The datasets we used in these experiments were randomly generated,
with all covariates drawn independently from uniform distributions on
the interval $[0,1]$, and responses then generated according to a
Gaussian process with specified hyperparameters.

We generated ten types of datasets in this way, with different combinations 
of the following:\vspace{-4pt}
        \begin{itemize}

                \item Number of
                observations: $n=300$ or $n=900$. \vspace{-4pt}

                \item Number of
                covariates: $p$=1 or $p=5$.  \vspace{-4pt}

                \item Type of covariance function: squared exponential
                covariance function with a single length
                scale (isotropic), or with multiple length scales 
                (Automatic Relevance Determination, ARD). Note that these
                are identical when $p=1$.\vspace{-4pt}

                \item Size of length scales: ``short'' indicates that
                a dataset has small length scales,``long'' that it has
                large length scales.\vspace{-4pt}

\end{itemize}
The specific hyperparameter values that were used for each combination of 
covariance function and length scale are shown in 
Table \ref{tbl:datasets}. 

        \begin{table}[b]
        \centering
        \begin{tabular}{c|c|c|c}
                \hline
                Length scale size                 &Length scale type & $\eta$& $l$\\ \hline
                short                                & isotropic        &5        & $l=0.1$ \\\hline
                short                                & ARD                &5        & $l_i=0.1i $ \\\hline
                long                                &  isotropic        &5        & $l=2$ \\\hline
                long                        &  ARD                &5        & $l_i=2i$ \\\hline

        \end{tabular}                
        \caption{Hyperparameter values used to generate the synthetic datasets.}
        \label{tbl:datasets}
        \end{table}

The efficiency of an MCMC method is usually measured by the
autocorrelation time, $\tau$, for the sequence of values produced by
the chain \citep[see][]{Neal:1993}:
        \begin{equation}
                \tau \ =\ 1 + 2\sum_{i=1}^{\infty} \rho_i
        \end{equation}
where $\rho_i$ is the lag-$i$ autocorrelation for some function of interest. 
In practice, with an
MCMC sample of size $M$, we can only find estimates, $\hat{\rho_i}$, of
autocorrelations up to lag $i=M-1$. To avoid excessive variance from
summing many noisy estimates, we typically estimate $\tau$ by
        \begin{equation}
                \hat{\tau} \ =\ 1 + 2\sum_{i=1}^{k} \hat{\rho_i}
        \end{equation}
where $k$ is a point where for all $i>k$, $\hat{\rho_i}$ is not 
significantly different from 0. 

Below, we will compare methods with respect to autocorrelation time of
the log likelihood.  For a fair comparison, we multiply the estimate
of each method's autocorrelation times by the average CPU time it
needs to obtain a new sample point.

\end{subsection}

\begin{subsection}{Experiments with mapping to a discretizing chain}

For each dataset, we tried the method of mapping to a discretizing
chain using both a $\pi^*$ formed with SOD and a $\pi^*$ formed with
Nystr\"om-Cholesky. For comparison, we also ran a standard MCMC model.
All the Markov chains were started from the hyperparameter values that
were used to generate them, so these tests assess only autocorrelation
time once the high-probability region of the posterior has been
reached, not time needed for convergence when starting at a
low-probability initial state.  The adjustable parameters of each
method were chosen to give good performance.  All chains were run for
2000 iterations, and autocorrelation times were then computed based on
the last two-thirds of the chain.

The standard MCMC method we used is a slice sampler \citep{Neal:2003},
specifically a univariate slice sampler with stepping-out and
shrinkage, updating parameters in sequence.  For the discretizing
Markov chain, the transition $R(x'|x)$ uses the same slice sampler.
Although slice sampling has tuning parameters (the stepsize, $w$, and
the upper limit on number of steps, $M$), satisfactory results can be
obtained without extensive tuning (that is, the autocorrelation time
of a moderately-well-tuned chain will not be much bigger than for an
optimally-tuned chain).  Because finding an optimal set of tuning
parameters is generally hard (requiring much time for trial runs), we
will accept the results using moderately-well-tuned chains.

We found that $r=s=1$ gives the best performance for the method of
mapping to a discretizing chain when the slice sampler is used for
$R(x'|x)$, at least if only fairly small values of $r$ and $s$ are
considered.  Recall that $r$ is the number of $\bar{T}$ updates to do
in each temporary mapping, and $s$ is the number of steps of $R(x'|x)$
to propose to move the mark for each $\bar{T}$ update.  Note that a
single slice sampling update will usually evaluate $\pi$ or $\pi^*$
more than once, since an evaluation is needed for each outward step
and each time a point is sampled from the interval found by stepping
out.  Therefore if we didn't use a mapping method we would have to
compute $\pi(x)$ several times for each slice sampling update. When a
mapping method is used, $\pi(x)$ only needs to be evaluated once each
update, for the new state (its value at the previous state having been saved),
while meanwhile, $\pi^*(x)$ will be evaluated several times.

We tuned the remaining parameter $m$, the subset size for SOD, or the
number of random columns for Nystr\"om-Cholesky, by trial and
error. Generally speaking, $m$ should be between 10\% and 50\% of $n$,
depending on the problem.  For Nystr\"om-Cholesky, quite good results
are obtained if such a value for $m$ makes $\pi^*$ be very close to
$\pi(x)$.


The results are in Table \ref{tbl:exp}, which shows CPU time per
iteration times autocorrelation time for the standard MCMC method, and
for other methods the ratio of this with the standard method. 
Table \ref{tbl:exp2} shows actual autocorrelation time and CPU time
per iteration for each experimental run.

        \begin{table}[p]
        \centering
        \footnotesize
        \begin{tabular}{c|c|c|c|c|c|c|r|r|c|c|c}
                \hline
        \multirow{2}{*}{\#} & \multicolumn{2}{c|}{Length scale} &\multirow{2}{*}{$p$} &\multirow{2}{*}{$n$}
                                & \multicolumn{3}{c|}{$m$} & \multicolumn{4}{|c}{$\!\!$Autocorrelation time $\times$ CPU time per iteration$\!\!$} \\ \cline{2-3}\cline{6-12}
        &size &type & & & SOD & NYS & TMP &$T_{\text{STD}}$ & $\!T_{\text{SOD}}/T_{\text{STD}}\!$ & $\!T_{\text{NYS}}/T_{\text{STD}}\!$ & $\!\!T_{\text{TMP}}/T_{\text{STD}}\!\!$ \\ \hline\hline
1	&	small	&	isotropic	&	1	&	300	&	40	&	30	&	40,\,20	&	0.76	&	0.45	&	0.51	&	1.05	\\ \hline
2	&	small	&	isotropic	&	5	&	300	&	150	&	-	&	100,\,50	&	1.62	&	0.81	&	-	&	0.14	\\ \hline
3	&	small	&	ARD	&	5	&	300	&	100	&	-	&	90,\,45	&	3.39	&	0.83	&	-	&	0.36	\\ \hline
4	&	long	&	isotropic	&	5	&	300	&	150	&	120	&	130,\,65	&	2.05	&	0.81	&	0.97	&	0.69	\\ \hline
5	&	long	&	ARD	&	5	&	300	&	90	&	80	&	100,\,50	&	5.23	&	0.66	&	0.85	&	0.51	\\ \hline
6	&	small	&	isotropic	&	1	&	900	&	60	&	90	&	60,\,30	&	9.06	&	0.27	&	0.23	&	0.28	\\ \hline
7	&	small	&	isotropic	&	5	&	900	&	300	&	-	&	-	&	18.17	&	0.51	&	-	&	-	\\ \hline
8	&	small	&	ARD	&	5	&	900	&	100	&	-	&	-	&	25.47	&	0.43	&	-	&	-	\\ \hline
9	&	long	&	isotropic	&	5	&	900	&	100	&	110	&	-	&	16.86	&	0.34	&	0.40	&	-	\\ \hline
10	&	long	&	ARD	&	5	&	900	&	300	&	90	&	-	&	47.46	&	0.67	&	0.34	&	-	\\ \hline

        \end{tabular}
        \caption{Results of experiments on the ten datasets.}
        \label{tbl:exp}
        \end{table}

        \begin{table}[p]
        \centering
        \footnotesize
        \begin{tabular}{c|c|c|r|r|r|r|r|r}
                \hline
                \multirow{2}{*}{\#} & \multicolumn{4}{c|}{CPU time (s) per iteration} & \multicolumn{4}{c}{Autocorrelation time} \\ \cline{2-9}
                 & STD & SOD & NYS & TMP & STD & SOD & NYS & TMP \\ \hline\hline
1	&	0.26	&	0.078	&	0.11	&	0.15	&	2.90	&	4.32	&	3.53	&	5.40	\\\hline
2	&	0.28	&	0.14	&	-	&	0.13	&	5.77	&	9.32	&	-	&	1.67	\\\hline
3	&	0.56	&	0.23	&	-	&	0.14	&	6.09	&	11.98	&	-	&	8.63	\\\hline
4	&	0.13	&	0.072	&	0.15	&	0.09	&	15.62	&	23.04	&	12.88	&	16.56	\\\hline
5	&	0.49	&	0.19	&	0.41	&	0.13	&	11.16	&	18.07	&	10.89	&	20.37	\\\hline
6	&	3.10	&	0.53	&	0.83	&	0.61	&	2.92	&	4.63	&	2.48	&	4.21	\\\hline
7	&	3.76	&	0.82	&	-	&	-	&	4.83	&	11.24	&	-	&	-	\\\hline
8	&	7.21	&	1.48	&	-	&	-	&	3.53	&	7.38	&	-	&	-	\\\hline
9	&	1.81	&	0.69	&	0.91	&	-	&	9.33	&	8.27	&	7.40	&	-	\\\hline
10	&	5.66	&	1.95	&	1.75	&	-	&	8.39	&	16.18	&	9.14	&	-	\\\hline

        \end{tabular}
        \caption{CPU time per iteration and autocorrelation time for each
                 run in Table \ref{tbl:exp}.}
        \label{tbl:exp2}
        \end{table}

From these results, we see that Subset of Data is overall the most
reliable method for forming a $\pi^*$. We can almost always find a SOD
type of $\pi^*$ that leads to more efficient MCMC than the standard
method.  Depending on the problem, mapping to a discretizing chain using
such a $\pi^*$ can be two to four times faster than standard MCMC, for
the Gaussian Process regression problems we tested.  The computational
savings go up when the size of the dataset increases. This is likely
because when $n$ is small, evaluation of $\pi$ is fast, so overhead
operations (especially those not related to $n$) are not trivial in
comparison.  The computational saving of $\pi^*$ compared to $\pi$
will be then less than the $m^3$ to $n^3$ ratio we expect from SOD for
large $n$.  Also when $n$ is small, time to compute $C$ (proportional
to $pn^2$) may be significant, which also reduces the computational
savings from a $\pi^*$ based on SOD.

For some datasets, we can find a Nystr\"om-Cholesky $\pi^*$ with a
small $m$ that can approximate $\pi$ well, in which case this method
works very nicely.  However, for datasets with small length scales
with $p=5$, in order to find a working $\pi^*$ we have to set $m$ to
be around $95\%$ of $n$ or greater, making $\pi^*$ as slow as, or even
slower than $\pi$.  This is due to the fact that when the length scale
parameters for the GP are small, the covariance declines rapidly as
the input variable changes, so $x$ and $x'$ that are even moderately
far apart have low covariance.  As a result, we were not able to find
efficient mapping method using Nystr\"om-Cholesky with performance
even close to standard MCMC (so no result is shown in the
table). On the other hand, when the length scale is large, a good
approximation can be had with a small $m$ (as small as $10\%$ of $n$).
For $n=900$ and $p=5$ with ARD covariance, Nystr\"om-Cholesky
substantially outperforms SOD.  

%

  \end{subsection}
%

\begin{subsection}{Experiments with tempered transitions}

We have seen in the previous section that the method of mapping to a
discretizing chain has a lot of tuning parameters, and finding the
optimal combination of these tuning parameters is not easy. The method
of tempered transitions actually has more tuning parameters. To start
with, we have to decide the number of ``layers'' (we call each of
$\hat{T}_i$ or $\check{T}_i$ a ``layer''). For each layer,
(e.g. $\hat{x}_i
\stackrel{\hat{T}_{i+1}}{\longrightarrow}\hat{x}_{i+1}$), we have to
decide how many MCMC updates to simulate.  This reduces the attraction
of tempered transitions, but in some situations it does improve
sampling efficiency.

In the experiments for the method of mapping to a discretizing chain,
the results given by both SOD and Nystr\"om-Cholesky for datasets with
$n=300, p=5$ are less satisfatory compared to others. We tried
tempered transitions with these datasets.  For simplicity, we used
just two layers, each of which uses SOD to form the transition. The
number of observations in each subset (denoted as $m_i$ for transition
$\hat{T}_i$ and $\check{T}_i$) is listed in Table \ref{tbl:exp} under
the column ``TMP'' and the time ratio results are under the column
``$T_\text{TMP}/T_\text{STD}$''.  We can see that for all these
datasets, tempered transitions outperform the method of mapping to a
discretizing chain, sometimes substantially.  The advantage of
tempered transitons is further illustrated n Figure
\ref{fig:comp_map_temper}, which shows the sample autocorrelation
plots of the log likelihood for both methods, on dataset \#2.

%

\begin{figure}[t]
        \centering

\noindent
\hfill \hspace{8pt}
  Mapping to a discretizing chain 
\hfill \hspace{38pt}
  Tempered transitions \hfill ~
\\[10pt]

            \includegraphics[scale=0.55,clip,bb=0in 0in 5.5in 5.6in]{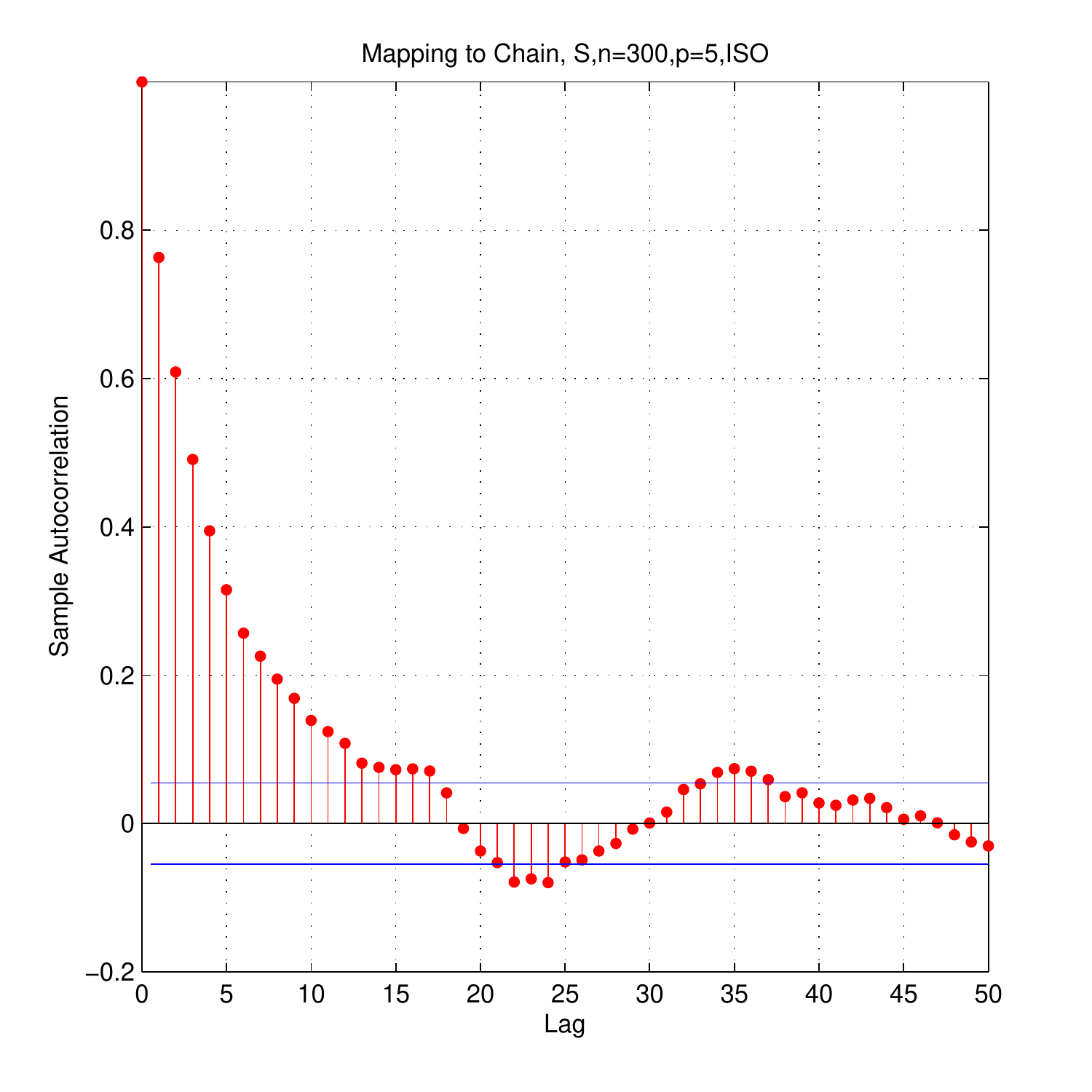}%
\hspace{14pt}%
            \includegraphics[scale=0.55,clip,bb=0in 0in 5.5in 5.6in]{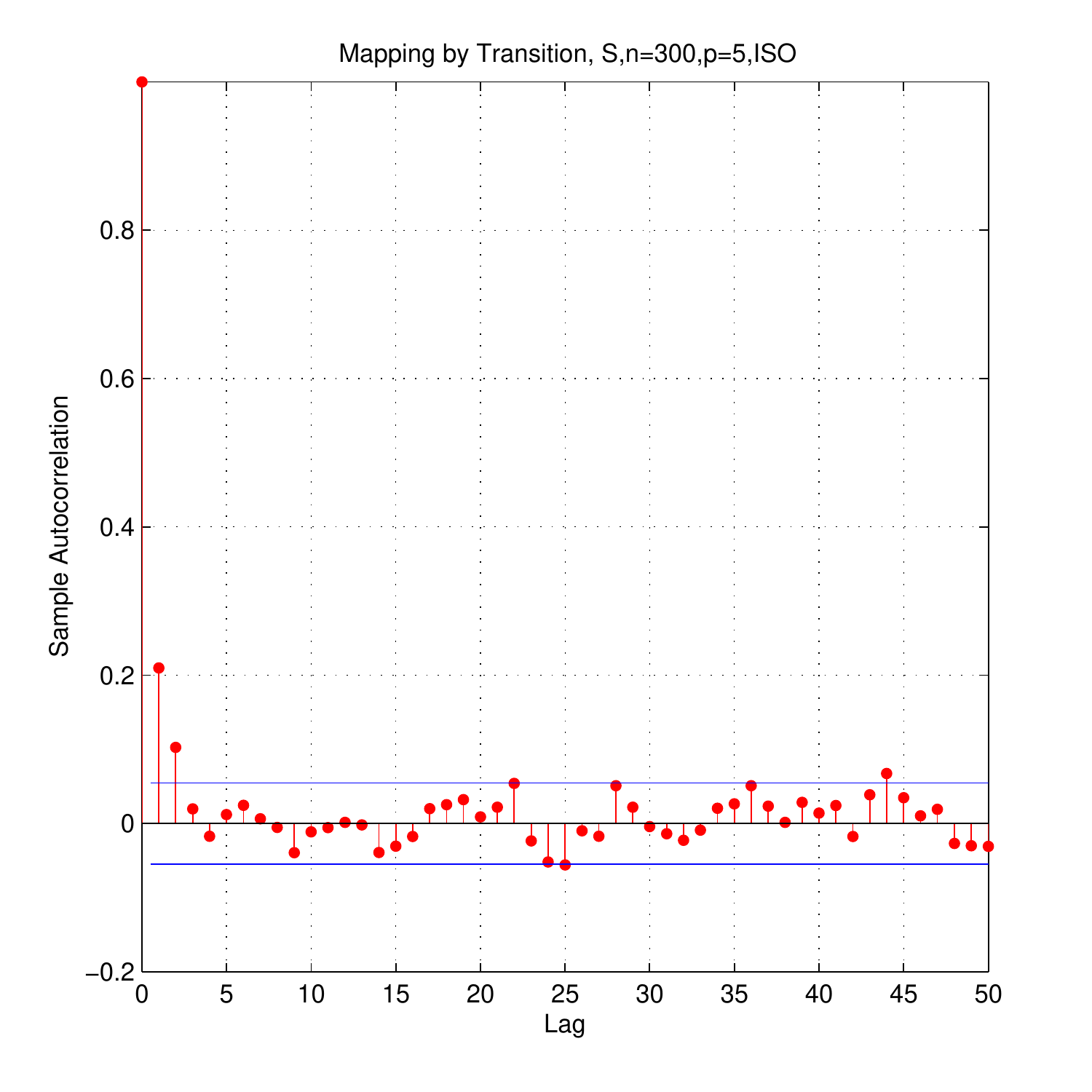}

\vspace{-20pt}

        \caption{Comparison of autocorrelation times of the log likelihood for
        MCMC runs using mapping to a
        discretizing chain and using tempered transitions.  Dataset \#2 is used
        (with five covariates, small length scales, an isotropic covariance 
        function, and 300 observations).}
        \label{fig:comp_map_temper}
\end{figure}

  \end{subsection}
 \end{section}

 \begin{section}{Discussion and future work}\label{sec:disc}

We have introduced two classes of MCMC methods using the ``mapping and
caching'' framework: the method of mapping to a discretizing chain,
and the tempered transition method. Our experiments indicate that for
method of mapping to a discretizing chain, when an appropriate $\pi^*$
is chosen (e.g. SOD approximation of $\pi$ with an appropriate $m$),
an efficient MCMC can be constructed by making ``local'' jumps (e.g.\
setting $r=s=1$). A good MCMC method can also be constructed using the
tempered transitions, with a small number of $\pi_i$, where each
$\hat{T}_i$ and $\check{T}_i$ makes only a small update.

These results are understandable.  Though $\pi^*$ and $\pi_i$, are
broader than $\pi$, making small adjustments a small number of times
will have a good chance to still stay in a high probability area of
$\pi$. However, even though the acceptance rate is high, this strategy
of making small adjustments cannot bring us very far from the previous
state. On the other hand, if we make large jumps, for instance, by
using large values for $r$ and $s$ in the method of mapping to a
discretizing chain, the acceptance rate will be low, but when a
proposal is accepted, it will be much further away from the previous
state, which is favourable for a MCMC method.  We haven't had much
success using this strategy so far, perhaps due to difficulty of
parameter tuning, but we believe this direction is worth pursuing.
The tempered transition method may be more suitable for this
direction, because moving from one state to another state further away
is somewhat similar to moving among modes --- the sequence of
$\hat{T}_i$ and $\check{T}_i$ should be able to ``guide'' the
transition back to a region with high probability under $\pi$.

\end{section}

\section*{Acknowledgements}

This research was supported by the Natural Sciences and Engineering Research
Council of Canada.  R.~N.\ holds a Canada Research Chair in Statistics and
Machine Learning.

\end{document}